\documentclass{amia}
\usepackage{lipsum} %Remove if not needed
\usepackage{tabularray}
\usepackage{xcolor}
\usepackage{longtable}
\usepackage{booktabs} % optional, for nicer rules
\usepackage{amsmath}
\usepackage{amssymb}

\begin{document}

\title{Contextual Phenotyping of Pediatric Sepsis Cohort Using Large Language Models 
}

\author{
    Aditya Nagori, PhD\textsuperscript{\rm 1,2},
    Ayush Gautam\textsuperscript{\rm 1,3},
    Matthew O. Wiens, PhD\textsuperscript{\rm 4,5,6},
    Vuong Nguyen, PhD\textsuperscript{\rm 4},\\
    Nathan Kenya Mugisha, MPH\textsuperscript{\rm 8},
    Jerome Kabakyenga, PhD\textsuperscript{\rm 9,10},
    Niranjan Kissoon, MD\textsuperscript{\rm 4,6,7},\\
    John Mark Ansermino, MD\textsuperscript{\rm 4,5,6},
    Rishikesan Kamaleswaran, PhD\textsuperscript{\rm 1, 2}
}

\institutes{
\small{
     $^1$ Department of Surgery, Duke university school of medicine, Durham NC, USA\\
 $^2$ Department of Anesthesiology, Duke university school of medicine, Durham NC, USA\\
   $^3$ Indian Institute of Technology, Goa, India\\
    $^4$Institute for Global Health, BC Children’s Hospital and BC Women’s Hospital + Health Centre, Vancouver, BC, Canada\\
     $^5$ Department of Anesthesia, Pharmacology \& Therapeutics, University of British Columbia, Vancouver, BC, Canada\\
     $^6$ BC Children’s Hospital Research Institute, BC Children’s Hospital, Vancouver, BC, Canada\\
    $^7$ Department of Pediatrics, University of British Columbia, Vancouver, BC, Canada\\
    $^8$ Walimu, Kampala, Uganda\\
     $^9$ Department of Community Health, Mbarara University of Science and Technology, Mbarara, Uganda\\
    $^10$ Maternal Newborn and Child Health Institute, Mbarara University of Science and Technology, Mbarara, Uganda
}
}

\maketitle

% Abstract
\begin{abstract}
Clustering patient subgroups is essential for personalized care and efficient resource use. Traditional clustering methods struggle with high-dimensional, heterogeneous healthcare data and lack contextual understanding. This study evaluates Large Language Model (LLM)-based clustering against classical methods using a pediatric sepsis dataset from a low-income country (LIC), containing 2,686 records with 28 numerical and 119 categorical variables. Patient records were serialized into text with and without a clustering objective. Embeddings were generated using quantized LLAMA 3.1 8B, DeepSeek-R1-Distill-Llama-8B  with low-rank adaptation(LoRA), and Stella-En-400M-V5 models. K-means clustering was applied to these embeddings. Classical comparisons included K-Medoids clustering on UMAP and FAMD-reduced mixed data. Silhouette scores and statistical tests evaluated cluster quality and distinctiveness. Stella-En-400M-V5 achieved the highest Silhouette Score (0.86). LLAMA 3.1 8B with the clustering objective performed better with higher number of clusters, identifying subgroups with distinct nutritional, clinical, and socioeconomic profiles. LLM-based methods outperformed classical techniques by capturing richer context and prioritizing key features. These results highlight the potential of LLMs for contextual phenotyping and informed decision-making in resource-limited settings.

\end{abstract}

\section{Introduction} Precision medicine aims to tailor healthcare by accounting for individual variability in genes, environment, and lifestyle \cite{collins2015new}. Key to this approach is identifying patient phenotypes that respond differently to treatments, thereby enabling targeted therapies \cite{ashley2016towards}. Unsupervised machine learning, particularly clustering algorithms, is instrumental in discovering these phenotypes within complex biomedical data \cite{kourou2015machine}. By grouping patients with similar clinical and molecular profiles, clustering supports tailored treatment plans and advances personalized care \cite{shah2012coming}.

Clustering in healthcare spans patient stratification by disease subtype to categorizing medical literature for evidence synthesis \cite{wang2018clinical}. However, classical algorithms like K-means and hierarchical clustering face challenges when applied to healthcare data. Medical datasets often include mixed numerical (e.g., lab results) and categorical (e.g., diagnosis codes) variables, and traditional methods usually require data transformation or dissimilarity measures that might not capture true relationships \cite{boriah2008similarity, ahmad2007k}. Moreover, high-dimensional data with numerous patient variables exacerbates the “curse of dimensionality,” diminishing the meaningfulness of distance metrics and increasing computational complexity \cite{bellman1961adaptive, aggarwal2001surprising, dean2008mapreduce}. Additionally, classical clustering may yield clusters that are hard to interpret for clinical decision-making \cite{doshi-velez2017towards}.

Large Language Models (LLMs), such as GPT-4, have shown impressive capabilities in understanding and generating human-like text \cite{brown2020language}. Their ability to process unstructured data and generate embeddings offers new opportunities for analyzing complex healthcare information \cite{devlin2019bert}. By converting mixed-type data into text and producing embeddings, LLMs can unify numerical and categorical data into a single space that captures inter-variable relationships \cite{zerveas2021transformer}. They also reduce dimensionality by distilling high-dimensional data into lower-dimensional representations that retain essential information \cite{mikolov2013efficient}. Furthermore, LLMs capture non-linear interactions often missed by classical methods and can enhance interpretability through explainable AI techniques \cite{vaswani2017attention, ribeiro2016trust}.

Despite these advantages, applying traditional clustering algorithms directly to LLM-generated embeddings in healthcare presents challenges \cite{esteva2019guide}. Issues include managing sensitive patient information, ensuring privacy, and maintaining clinical interpretability. Transparent clustering is vital for clinical translation; clinicians must understand what distinguishes patient subgroups based on clinical features or trajectories. Classical methods often act as “black boxes” without clear explanations, limiting adoption where clusters must be both clinically meaningful and statistically robust.

To address these gaps, we propose a novel pipeline that leverages LLM-based dynamic embedding generation, context-aware dimensionality reduction, and phenotyping algorithms. Unlike static embeddings, our approach emphasizes features relevant to clustering outcomes. We demonstrate the feasibility of our pipeline using a pediatric sepsis cohort—a compelling case given its clinical heterogeneity, temporal complexity, and need for early intervention.

Our pipeline serializes mixed-type EHR data into text, generates embeddings with LLMs, and employs dimensionality reduction to preserve semantic structure. We then evaluate clustering performance and the clinical coherence of the resulting phenotypes against traditional methods, advancing the integration of foundation models into clinical data science.

\section{Methods}

\subsection{Study Population and Data Collection}
We utilized a synthetic dataset \cite{huxford2024pediatric} based on a prospective, multisite observational cohort study of children aged 6 to 60 months with suspected sepsis admitted to hospitals in Uganda between 2017 and 2020 \cite{wiens2023mortality} The synthetic dataset was generated using the \texttt{synthpop} package in R, applying the non-parametric classification and regression tree (CART) method for variable synthesis \cite{nowok2016synthpop}. The dataset included demographics, vital signs, lab values, symptoms, comorbidities, medications, socio-environmental factors, and outcomes such as mortality and length of stay.  

\subsection{Study Design}
This secondary analysis aimed to phenotype pediatric sepsis patients using advanced machine learning techniques, including transformer-based models and clustering methods.

\subsection{Data Serialization and Embedding Generation Using Language Models}
Let \( X \) be a dataset with \( n \) samples (patient records) and \( m \) features (variables). Each patient record \( x_i \) is represented as:
\[
x_i = (x_{i1}, x_{i2}, \dots, x_{im}),
\]
where \( x_{ij} \) represents the value of the \( j \)-th feature for the \( i \)-th patient.

To serialize the data, each record \( x_i \) is converted into a serialized text format \( s_i \):
\[
s_i = \text{serialize}(x_i) = \text{concatenate}(x_{i1}, x_{i2}, \dots, x_{im}).
\]
\subsection{Embedding Generation Using Language Models}
We utilized three advanced language models to generate embeddings from serialized patient data. The first two models, the Quantized LLama 3.1 8B model \cite{touvron2024llama} and the DeepSeek-R1-Distill-Llama-8B model \cite{turn0search0, deepseek2025r1}, were employed with 4-bit quantization and accessed via the \texttt{unsloth} Python library with a LoRA \cite{hu2021lora} adapter. Let the serialized patient record for the \(i\)-th patient be denoted as \(s_i\); the corresponding embedding is generated as:
\[
e_{i,\text{llm}} = f_{\text{llm}}(s_i; \theta_{\text{llm}}),
\]
where \(f_{\text{llm}}(\cdot; \theta_{\text{llm}})\) represents either the LLama 3.1 8B or DeepSeek-R1-Distill-Llama-8B model, and \(e_{i,\text{llm}} \in \mathbb{R}^{4096}\) is the high-dimensional embedding for the \(i\)-th patient. Quantization improves computational efficiency, while the LoRA adapter minimizes the number of parameters updated during fine-tuning.

The third model, Stella-en-400M-v5, was accessed via the \texttt{sentence\_transformers} Python library. Denoting its function as \(f_{\text{Stella}}(\cdot; \theta_{\text{Stella}})\), the embedding for the \(i\)-th patient is generated as:
\[
e_{i,\text{Stella}} = f_{\text{Stella}}(s_i; \theta_{\text{Stella}}),
\]
with \(e_{i,\text{Stella}} \in \mathbb{R}^{1024}\). Together, these models provide complementary embeddings that capture both task-specific and general semantic features of the patient data.

\subsection{Incorporation of Clustering Objective}
To prioritize critical features for clustering, we appended the clustering objective to each serialized patient record. The clustering objective was: "Generate an embedding for clustering patients based on their physiological severity and prioritize features indicative of critical conditions." This objective guided the language model to generate embeddings tailored for clustering tasks. Formally, let the serialized patient record for the \( i \)-th patient be \( s_i \), and let the clustering objective be represented as \( O \). The modified serialized record \( s_i' \) is:
\[
s_i' = \text{concatenate}(s_i, O).
\]

The sentence transformer model \( f(\cdot; \theta) \), parameterized by \( \theta \), generates the embedding \( e_i' \) for the modified record \( s_i' \):
\[
e_i' = f(s_i'; \theta).
\]

Here:
\begin{itemize}
    \item \( s_i \) is the original serialized text of patient data.
    \item \( O \) is the clustering objective appended to the serialized text.
    \item \( f(\cdot; \theta) \) is the language model that generates embeddings.
    \item \( e_i' \in \mathbb{R}^d \) is the resulting task-specific embedding in a \( d \)-dimensional space.
\end{itemize}

By incorporating \( O \) into \( s_i \), the embedding process explicitly emphasizes features critical for clustering tasks, aligning the generated embeddings with clustering objectives.

\subsection{Classical Clustering Approaches for Comparison}
\subsubsection{Clustering on Mixed Data UMAP Embeddings}
We applied UMAP (Uniform Manifold Approximation and Projection) to reduce the dimensionality of numerical features while preserving global structure \cite{mcinnes2018umap}. For categorical features, UMAP was applied using the Dice metric. The numeric and categorical UMAP embeddings were concatenated and K-Medoids clustering was performed with Euclidean distance. Silhouette scores were computed for cluster numbers ranging from 2 to 9.

\subsubsection{Clustering on FAMD Embeddings}
Factor Analysis of Mixed Data (FAMD) from the Python \texttt{prince} library \cite{halford2024prince} was used to reduce dimensionality while handling mixed data types \cite{pages2014mfa}. K-Medoids clustering was applied, and silhouette scores were computed for 2 to 9 clusters.

\subsection{Descriptive Statistics and Statistical Testing}
\subsubsection{Descriptive Statistics}
Median and interquartile range (IQR) were calculated for continuous variables (e.g., age, height, weight, MUAC, heart rate). Frequencies and percentages were calculated for categorical variables (e.g., symptoms, comorbidities, vaccination status).

\subsubsection{Statistical Testing}
We performed the Kruskal-Wallis test \cite{kruskal1952use} to compare continuous variables and the chi-square test to compare categorical variables between clusters, applying a Bonferroni correction to control for multiple comparisons and determine the statistical significance of observed differences.

\section{Results}

\subsubsection{Performance Comparison of Clustering Algorithms}

The clustering performance of various algorithms was assessed using Silhouette Scores across different numbers of clusters. The algorithms evaluated included Stella-en-400M-v5, Llama 3.1 8b, DeepSeek-R1-Distil-Llama-8b and their variants with clustering objectives incorporated, and K-Medoids clustering applied on UMAP and FAMD embeddings.

\begin{figure}[h]
  \centering
  \includegraphics[width=\linewidth]{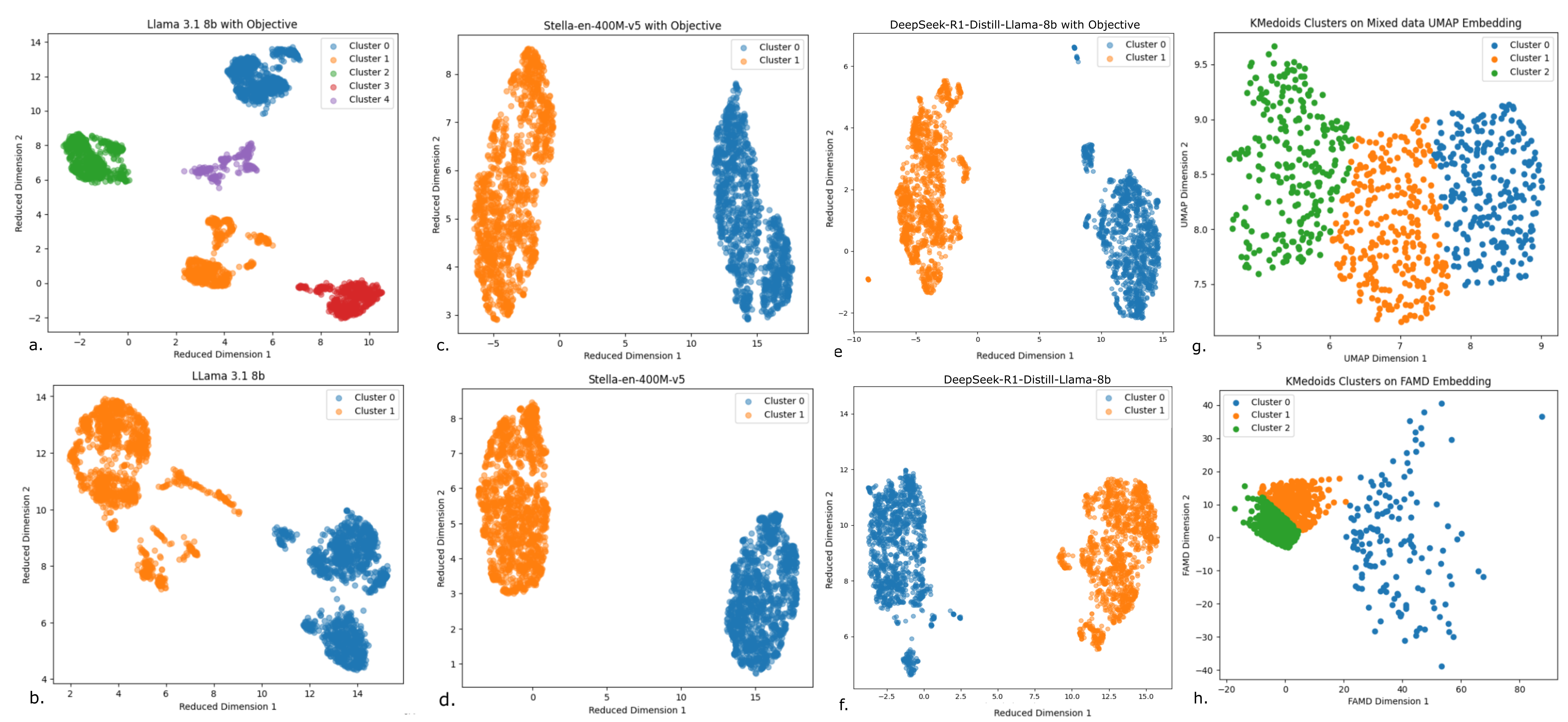}
  \caption{Clustering Performance Metrics Across Different Models Fig 1a). Clusters on Llama 3.1 8b model embeddings with objectives inserted in the serialized data. Fig 1b). Clusters on Llama 3.1 8b model embeddings. Fig 1c). Clusters on Stella model embeddings with objectives inserted in the serialized data.  Fig 1d). Clusters on Stella model embeddings. Fig 1e). Clusters on DeepSeek-R1-Distill-Llama-8b model embeddings with objectives inserted in the serialized data. Fig 1f). Clusters on DeepSeek-R1-Distill-Llama-8b model embeddings. Fig 1g.) Kmedoid clusters on mixed data Umap Embeddings. Fig 1h.) Kmedoid clusters on FAMD Embeddings. } 
  \label{fig:clusters}
\end{figure}

\begin{figure}[h]
  \centering
  \includegraphics[width=0.5\textwidth]{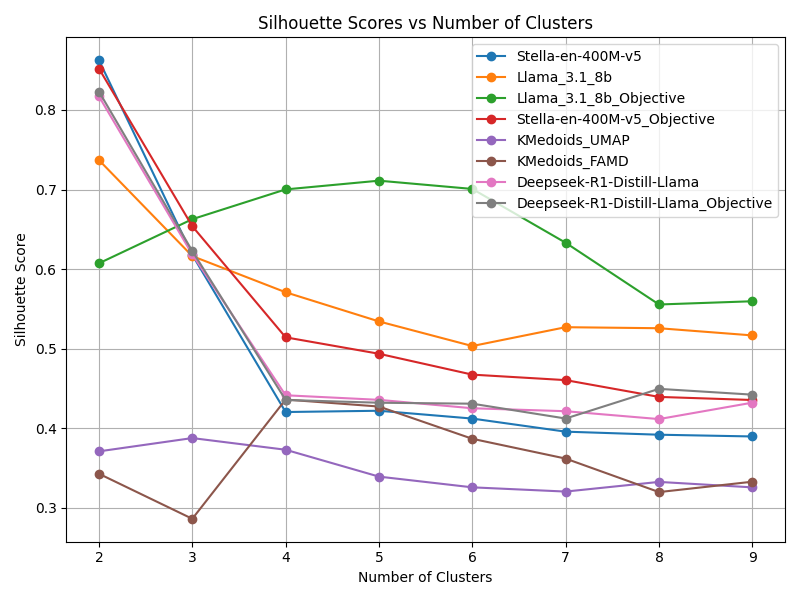}
  
  \caption{ Fig 2). Silhouette score vs number of clusters}
  \label{fig:scores}
\end{figure}

Llama 3.1 8b performed well, with a Silhouette Score of 0.74 at two clusters (Fig. \ref{fig:clusters}c, Fig. \ref{fig:scores}) , decreasing to 0.52 at nine clusters. When incorporating the clustering objective (Llama 3.1 8b with Objective), the algorithm showed improved performance at higher cluster numbers  (Fig. \ref{fig:clusters}a) (Fig. \ref{fig:scores}). The Silhouette Score peaked at 0.71 with five clusters   (Fig. \ref{fig:scores}).
Stella-en-400M-v5 achieved the highest Silhouette Score of 0.86 at two clusters (Fig. \ref{fig:clusters}d, Fig. \ref{fig:scores}), indicating well-defined and cohesive clustering. However, the Silhouette Score decreased as the number of clusters increased, reaching 0.39 at nine clusters  (Fig. \ref{fig:scores}).

Similarly, Stella-en-400M-v5 with a clustering objective maintained  (Fig. \ref{fig:clusters}c)  high Silhouette Scores across cluster numbers, starting at 0.85 for two clusters and decreasing to 0.44 at nine clusters (Fig. \ref{fig:scores}). While slightly lower than its non-objective variant at two clusters, it demonstrated better performance at higher cluster counts, suggesting that incorporating clustering objectives helps maintain cluster integrity as complexity increases.

The K-Medoids clustering on UMAP embeddings (Fig. \ref{fig:clusters}g)  resulted in moderate Silhouette Scores ranging from 0.37 to 0.34 across two to nine clusters (Fig. \ref{fig:scores}). On FAMD embeddings (Fig. \ref{fig:clusters}h), K-Medoids showed slightly better results, with the highest Silhouette Score of 0.44 at four clusters (Fig. \ref{fig:scores}), suggesting improved cluster compactness and separation at this cluster number.

Overall, the LLM-based embeddings, particularly when combined with clustering objectives, performed better than the K-Medoids methods on both UMAP and FAMD embeddings (Fig. \ref{fig:clusters}g, \ref{fig:clusters}h). The incorporation of clustering objectives into the embedding generation process enhanced the ability of the models to produce task-specific embeddings, leading to better-defined clusters. Stella-en-400M-v5 and Llama 3.1 8b with objective demonstrated superior clustering performance, with the former excelling at lower cluster numbers Fig. \ref{fig:clusters}c) and the latter showing optimal results at five clusters Fig. \ref{fig:clusters}a).

\subsubsection{Descriptive Statistics and Group Comparisons }

The clustering analysis reveals five distinct clusters using LLama 3.1 8b model embeddings (Cluster 0 to Cluster 4) characterized by various demographic, clinical, and socioeconomic variables. Statistical significance among clusters was assessed using the Kruskal-Wallis test for continuous variables and the Chi-Square test for categorical variables.

Cluster 0 includes children with a median (IQR) age at admission of 19.05 (13.3) months (p-value = 0). The growth z-scores i.e. Weight-for-Length (WFL) z-score: -1.03 (2.27), p-value = 0, Body Mass Index (BMI) z-score: -0.93 (2.46), p-value = 0, Weight-for-Age (WFA) z-score: -1.28 (2.08), p-value = 0, indicating slightly underweight children for their age and height. A notable characteristic is the high usage of intravenous ampicillin or amoxicillin antibiotics, with 99.73\% of children receiving these medications (p-value = 0). Families in Cluster 0 have moderate access to municipal water (46.29\%, p-value = 0), and a higher percentage use water purification methods (80.49\%, p-value = 0), indicating slightly better socioeconomic conditions. The median hematocrit level is higher at 36\% (p-value = 0), which may reflect better overall health. Vital signs are stable, with a median (IQR) heart rate of 142 (29) bpm and respiratory rate of 44 (20) breaths per minute (both p-value = 0). 

Cluster 1 consists of younger children with a median age at admission of 12.10 (13.9) months (p-value = 0). They have lower growth z-scores i.e. WFL z-score (-0.99 (2.7), p-value = 0), BMI z-score (-0.98 (2.65), p-value = 0), WFA z-score (-1.13 (2.31), p-value =0), possibly indicating moderately poor nutritional status. The median (IQR) heart rate is slightly higher at 149 (31) bpm (p-value = 0), and the respiratory rate is slightly elevated at 47 (20) breaths per minute (p-value = 0). This cluster has the highest in-hospital mortality rate at 6.56\% (p-value = 0.001). Clinical symptoms such as severe respiratory distress are significantly more prevalent (24.29\%, p-value = 0), and there is a higher incidence of coma (4.96\%, p-value = 0). Measles vaccination rates are significantly lower, with only 24.82\% vaccinated (p-value = 0), indicating a gap in preventive healthcare. Mothers in Cluster 1 have lower education levels (p-value = 0.0001), which may contribute to poorer health outcomes in children.

Cluster 2 comprises children with the highest median age at admission, 21.30 months (p-value = 0), indicating they are the oldest among all clusters. Their growth z-scores i.e. WFA z-score: -0.99 (1.84), p-value = 0, BMI z-score: -0.81 (2.2), p-value = 0, and WFL z-score: -0.89 (2.1), p-value = 0 are above negative 1 standard deviation, reflecting possible better nutritional status. The median heart rate is lower at 139 (28) bpm (p-value = 0), and respiratory rate is the lowest among clusters at 43 (17) breaths per minute (p-value = 0), suggesting more stable vital signs. Significantly, Cluster 2 has the highest percentage of malaria-positive cases (36.42\%, p-value = 0). The primary water source for families in this cluster is boreholes (31.56\%, p-value = 0). Additionally, children in this cluster have longer durations of exclusive breastfeeding, with 50.38\% exclusively breastfed for 6 months (p-value = 0), indicating better early-life nutrition practices.

Cluster 3 includes the youngest children with a median age at admission of 10.45 months (p-value = 0). Nutritional indicators are the poorest among all clusters, with all growth z-scores below negative 1 standard deviation i.e. WFL z-score: -1.26 (2.23), p-value = 0, BMI z-score: -1.31 (2.38), p-value=0, WFA z-score: -1.55 (2.50), p-value =0. The cluster exhibits higher rates of prolonged diarrhea (41.57\%, p-value = 0.0008) and symptoms like rash (23.37\%, p-value = 0), indicating more severe and perhaps chronic health issues.

Mothers in Cluster 3 have the lowest maternal education levels (p-value = 0.0001) and higher rates of unknown age at first pregnancy (p-value = 0). There is also a significant percentage of premature births (4.41\%, p-value = 0.026) and low birth weight infants (4.60\%, p-value = 0.0025). These factors contribute to the overall vulnerability of children in this cluster.

Cluster 4 has children with a median age at admission of 18.40 months (p-value = 0), similar to Cluster 0. Nutritional status is moderate, with growth z-scores WFL z-score: -0.96 (2.19), p-value = 0, BMI z-score: -0.93 (2.20), p-value =0, and WFA z-score: -1.13 (1.92), p-value = 0. A distinctive feature of Cluster 4 is a higher incidence of symptoms, such as changes in urine color (19.72\%, p-value = 0.0001). Additionally, there is a slightly higher usage of HIV medications (0.94\%, p-value = 0.0039).
Families in Cluster 4 have moderate access to municipal water (44.13\%, p-value = 0) and water purification methods (75.59\%, p-value = 0). The median respiratory rate is the lowest among clusters at 43 (20) breaths per minute (p-value = 0), indicating stable vital signs.

\scriptsize
\begin{longtable}{| l | c | c | c | c | c | c |}
\caption{Cluster-level summary of clinical, laboratory, and socioeconomic characteristics in a pediatric sepsis cohort. Data for each variable are presented across five patient clusters with corresponding p-values indicating statistical significance.} \label{tab:multi}\\
\hline
\textbf{Variable} & \textbf{Cluster 0} & \textbf{Cluster 1} & \textbf{Cluster 2} & \textbf{Cluster 3} & \textbf{Cluster 4} & \textbf{p-value} \\
\hline
\endfirsthead

\multicolumn{7}{c}{\tablename\ \thetable{} -- continued from previous page} \\
\hline
\textbf{Variable} & \textbf{Cluster 0} & \textbf{Cluster 1} & \textbf{Cluster 2} & \textbf{Cluster 3} & \textbf{Cluster 4} & \textbf{p-value} \\
\hline
\endhead

\hline \multicolumn{7}{r}{Continued on next page}\\
\endfoot

\hline
\endlastfoot

\multicolumn{7}{l}{\textbf{Anthropometrics}}\\
Age at Admission (months) & 19.05 (13.30) & 12.10 (13.90) & 21.30 (16.75) & 10.45 (10.65) & 18.40 (15.60) & 0\\
Height (cm)               & 78.90 (11.62) & 73.35 (14.50) & 82.00 (13.00) & 71.90 (11.20) & 80.00 (15.00) & 0\\
Weight (kg)               & 9.50 (3.00)   & 8.70 (2.80)   & 10.00 (2.65)  & 7.80 (2.79)   & 9.50 (3.00)   & 0\\
MUAC (mm)                 & 140.00 (18.00)& 137.00 (21.00)& 142.00 (18.00)& 134.00 (20.00)& 140.00 (16.00)& 0\\

\multicolumn{7}{l}{\textbf{Vital Signs}}\\
Heart Rate (bpm)          & 142.00 (29.00)& 149.00 (31.00)& 139.00 (28.00)& 146.50 (33.00)& 143.00 (27.00)& 0\\
Respiratory Rate (breaths/min) & 44.00 (20.00) & 47.00 (20.00) & 43.00 (17.00) & 47.50 (19.25) & 43.00 (20.00) & 0\\
Hematocrit (\%)           & 36.00 (10.00)& 32.00 (13.75)& 32.00 (16.00)& 34.00 (8.00)& 34.00 (10.00)& 0\\

\multicolumn{7}{l}{\textbf{Oxygen Measures}}\\
Oxygen Saturation (1st measure) & 97.00 (5.00) & 97.00 (7.00) & 97.00 (4.75) & 97.00 (8.00) & 97.00 (5.00) & 1e-04\\
Oxygenation Index (sqi2\_perc\_oxi\_adm) & 88.00 (23.75)& 84.50 (33.25)& 89.00 (27.00)& 85.00 (31.00)& 86.00 (30.25)& 2e-04\\
SpO\textsubscript{2} Measured on Oxygen   & 9.07\%       & 25.35\%       & 10.02\%       & 14.94\%       & 12.21\%       & 0\\
O\textsubscript{2} available \& used      & 14.84\%      & 29.43\%       & 14.57\%       & 22.61\%       & 15.96\%       & 0\\
O\textsubscript{2} available \& not used  & 80.91\%      & 65.78\%       & 81.34\%       & 72.99\%       & 78.87\%       & 0\\
Severe respiratory distress               & 10.99\%      & 24.29\%       & 9.56\%        & 19.16\%       & 10.33\%       & 0\\
Capillary refill time                     & 7.55\%       & 14.72\%       & 14.57\%       & 11.69\%       & 10.33\%       & 2e-04\\

\multicolumn{7}{l}{\textbf{Blantyre Coma Scale (BCS)}}\\
BCS Eye: fails to watch/follow        & 0.96\% & 22.87\% & 0.91\% & 1.15\% & 6.10\% & 0\\
BCS Eye: watches/follows              & 99.04\%& 77.13\% & 99.09\%& 98.85\%& 93.90\%& 0\\
BCS Motor: localizes pain             & 97.39\%& 73.40\% & 96.66\%& 95.40\%& 93.43\%& 0\\
BCS Motor: no/inappropriate response  & 0.00\% & 6.21\%  & 0.15\% & 0.00\% & 0.47\% & 0\\
BCS Motor: withdraws from pain        & 2.61\% & 20.39\% & 3.19\% & 4.60\% & 6.10\% & 0\\
BCS Verbal: appropriate cry/speech    & 99.31\%& 74.47\% & 99.54\%& 99.04\%& 96.24\%& 0\\
BCS Verbal: moan/abnormal cry         & 0.69\% & 18.97\% & 0.46\% & 0.96\% & 3.76\% & 0\\
BCS Verbal: no response               & 0.00\% & 6.56\%  & 0.00\% & 0.00\% & 0.00\% & 0\\

\multicolumn{7}{l}{\textbf{Antibiotics on Admission}}\\
IV Ampicillin/Amoxicillin  & 99.73\% & 13.83\% & 0.00\% & 94.83\% & 51.64\% & 0\\
PO Ampicillin/Amoxicillin  & 0.14\%  & 3.90\%  & 1.97\% & 1.34\%  & 0.00\%  & 0\\
IV Penicillin              & 0.41\%  & 24.29\% & 25.64\%& 0.77\%  & 15.02\% & 0\\
IV Ceftriaxone             & 3.98\%  & 55.50\% & 63.28\%& 4.60\%  & 30.99\% & 0\\
IV Gentamicin              & 91.07\% & 56.56\% & 47.04\%& 92.34\% & 66.67\% & 0\\
IV/PO Antimalarial         & 18.68\% & 30.32\% & 33.84\%& 15.52\% & 19.72\% & 0\\

\multicolumn{7}{l}{\textbf{Vaccination Status}}\\
Measles vacc unknown & 0.00\% & 0.35\%  & 0.00\% & 10.73\% & 2.35\%  & 0\\
Measles vacc yes     & 100.00\%&24.82\% & 99.09\%& 4.02\%  & 65.26\% & 0\\
Source measles vacc (none)      & 0.00\% & 75.18\% & 0.91\%  & 95.98\% & 34.74\% & 0\\
Source measles vacc: card       & 16.62\%& 3.90\%  & 16.54\% & 0.57\%  & 8.92\%  & 0\\
Source measles vacc: self-report& 83.38\%& 20.92\% & 82.55\% & 3.45\%  & 56.34\% & 0\\
Pneumo vacc 0 doses             & 0.27\% & 8.51\%  & 0.61\%  & 6.90\%  & 1.88\%  & 0\\
Pneumo vacc 1 dose              & 0.27\% & 5.32\%  & 0.61\%  & 6.13\%  & 3.29\%  & 0\\
Pneumo vacc 2 doses             & 7.42\% & 14.89\% & 5.77\%  & 13.22\% & 11.27\% & 0\\
Pneumo vacc 3 doses             & 92.03\%& 68.97\% & 93.02\% & 60.34\% & 79.81\% & 0\\
Pneumo vacc unknown             & 0.00\% & 2.30\%  & 0.00\%  & 13.41\% & 3.76\%  & 0\\
Source pneumo vacc (none)       & 0.00\% & 10.64\% & 0.15\%  & 20.50\% & 5.63\%  & 0\\
Source pneumo vacc: self-report & 83.38\%& 69.50\% & 83.61\% & 63.22\% & 81.22\% & 0\\
DPT/Penta vacc 0 doses          & 0.27\% & 4.79\%  & 0.46\%  & 6.13\%  & 1.41\%  & 0\\
DPT/Penta vacc 1 dose           & 0.27\% & 5.32\%  & 0.30\%  & 5.17\%  & 2.82\%  & 0\\
DPT/Penta vacc 2 doses          & 7.42\% & 15.25\% & 5.61\%  & 13.22\% & 11.27\% & 0\\
DPT/Penta vacc 3 doses          & 92.03\%& 73.05\% & 93.63\% & 63.03\% & 81.22\% & 0\\
DPT/Penta vacc unknown          & 0.00\% & 1.60\%  & 0.00\%  & 12.45\% & 3.29\%  & 0\\
Source DPT/Penta vacc (none)    & 0.00\% & 6.21\%  & 0.00\%  & 18.20\% & 4.69\%  & 0\\
Source DPT/Penta vacc: self-report & 83.79\%& 74.11\% & 83.76\% & 65.52\% & 81.69\% & 0\\

\multicolumn{7}{l}{\textbf{Recent Medication Usage}}\\
Antibiotics used in past week: don't know & 0.00\% & 6.38\% & 0.00\% & 0.00\% & 0.00\% & 0\\
Antibiotics used in past week: no        & 51.10\%& 41.31\%& 51.90\%& 42.72\%& 45.07\%& 1e-04\\
Antimalarials used in past week: don't know & 0.00\% & 5.67\% & 0.00\% & 0.00\% & 0.00\% & 0\\
Antimalarials used in past week: no         & 72.25\%& 58.33\%& 65.10\%& 69.16\%& 63.85\%& 0\\

\multicolumn{7}{l}{\textbf{Symptoms}}\\
Rash             & 7.28\%  & 15.60\% & 7.44\%  & 23.37\% & 10.80\% & 0\\
Urine color changes & 10.71\% & 17.02\% & 18.36\% & 12.84\% & 19.72\% & 1e-04\\
Coma             & 0.14\%  & 4.96\%  & 0.30\%  & 0.38\%  & 0.00\%  & 0\\
Co-morbid dx unknown & 1.65\% & 5.32\%  & 1.21\%  & 6.51\%  & 1.41\%  & 0\\

\multicolumn{7}{l}{\textbf{Feeding/Breastfeeding}}\\
Feeding poorly   & 77.75\% & 62.23\% & 75.72\% & 75.67\% & 74.65\% & 0\\
Not feeding      & 3.57\%  & 21.63\% & 5.16\%  & 4.98\%  & 4.69\%  & 0\\
No cough/choke with liquids & 47.80\% & 38.30\% & 44.16\% & 38.12\% & 51.17\% & 1e-04\\
6 months exclusive BF   & 54.40\% & 45.21\% & 50.38\% & 40.80\% & 51.17\% & 0\\
Unknown exclusive BF     & 1.10\%  & 1.42\%  & 1.21\%  & 7.28\%  & 2.82\%  & 0\\
Total BF \textgreater 12 months & 41.48\% & 24.11\% & 48.71\% & 19.35\% & 38.50\% & 0\\
Still being breastfed    & 38.19\% & 59.57\% & 30.35\% & 60.34\% & 40.38\% & 0\\
Unknown BF duration      & 0.69\%  & 0.71\%  & 1.06\%  & 3.45\%  & 0.94\%  & 2e-04\\

\multicolumn{7}{l}{\textbf{Travel}}\\
Motorcycle travel        & 55.77\% & 42.02\% & 41.43\% & 53.45\% & 49.77\% & 0\\
Private vehicle travel   & 1.92\%  & 2.66\%  & 5.92\%  & 1.92\%  & 5.63\%  & 0\\
Taxi/special hire travel & 39.01\% & 49.29\% & 48.56\% & 40.61\% & 42.25\% & 2e-04\\

\multicolumn{7}{l}{\textbf{Other Clinical/Maternal Info}}\\
Good health prior to illness & 87.64\% & 83.69\% & 86.49\% & 80.08\% & 92.49\% & 0\\
Mother's age known          & 97.53\% & 97.34\% & 96.36\% & 91.00\% & 95.31\% & 0\\
Mother's age at first pregnancy known & 95.19\% & 91.31\% & 92.26\% & 84.48\% & 92.02\% & 0\\
Mother's education unknown  & 0.69\%  & 1.24\%  & 0.00\%  & 2.87\%  & 0.94\%  & 1e-04\\
Maternal HIV unknown        & 4.95\%  & 5.32\%  & 5.92\%  & 11.69\% & 7.51\%  & 0\\

\multicolumn{7}{l}{\textbf{Household/Environment}}\\
Borehole water source       & 18.96\% & 28.37\% & 31.56\% & 19.54\% & 22.07\% & 0\\
Uses water purification     & 80.49\% & 61.35\% & 64.34\% & 74.52\% & 75.59\% & 0\\

\multicolumn{7}{l}{\textbf{Malaria}}\\
Malaria test positive       & 25.14\% & 31.74\% & 36.42\% & 20.50\% & 26.76\% & 0\\

\multicolumn{7}{l}{\textbf{Admission/Nutritional Indices}}\\
Length of admission (days)  & 4.00 (3.00) & 4.00 (4.00) & 4.00 (4.00) & 5.00 (4.75) & 4.00 (4.00) & 0\\
Weight-for-length Z-score   & -1.03 (2.27)& -0.99 (2.70)& -0.89 (2.10)& -1.26 (2.23)& -0.96 (2.19)& 0\\
BMI Z-score                 & -0.93 (2.46)& -0.98 (2.65)& -0.81 (2.20)& -1.31 (2.38)& -0.93 (2.20)& 0\\
Weight-for-age Z-score      & -1.28 (2.08)& -1.13 (2.31)& -0.99 (1.84)& -1.55 (2.50)& -1.13 (1.92)& 0\\

% Continue with additional rows and sections as needed...
\end{longtable}
\normalsize  % Ensure default font size resumes
\section{Discussion}
The use of Large Language Models (LLMs), particularly the LLAMA 3.1 8B model, in clustering structured healthcare data represents a significant advance over traditional methods that struggle with high-dimensional, heterogeneous datasets \cite{aggarwal2001surprising}. LLMs leverage deep architectures and non-linear activations to capture complex patterns and relationships \cite{naveed2024overview, ester1996density, lecun2015deep}, which is crucial for healthcare data with mixed types and intricate interactions.

Our pipeline incorporated the dynamic embedding generation by incorporation of clustering objectives. We embedded clustering objectives into the prompts used for generating embeddings, ensuring that the embeddings are task-specific and capture relevant features for clustering. We introduced the task-Specific embeddings with LLMs. By utilizing an LLM, we generated embeddings that unify mixed data types and capture complex relationships, addressing challenges with traditional methods. We resolved the scalability and efficiency issue by implementing mini-batch processing and parallel computing to handle large datasets efficiently, addressing computational challenges.

LLMs are pre-trained on vast amounts of data, allowing them to incorporate contextual understanding into their analysis\cite{shah2024large}. We used a quantized Llama 3.1 8B and DeepSeek-R1-Distill-Llama-8b Models with LoRA Adapter, the these model were chosen for its open-access weights and state-of-the-art performance. We applied quantization techniques to reduce the model size and computational requirements. Low-Rank Adaptation (LoRA) was used to fine-tune the model efficiently on our dataset without updating all model parameters \cite{hu2021lora}. The model generated embeddings for each serialized record with and without the clustering objective appended. We also used the Stella-En-400M-V5 Model, the Stella-En-400M-V5 model was selected due to its high performance on the Massive Text Embedding Benchmark (MTEB) leaderboard and smaller model size. Importantly, it is trained based on Alibaba-NLP/gte-large-en-v1.5 and Alibaba-NLP/gte-Qwen2-1.5B-instruct which is trained on 75 languages, including Ugandan languages, making it well-suited for our dataset \cite{zhang2024mgte}. Like the LLAMA model, embeddings were generated with and without the clustering objective. Although our dataset is structured, the LLAMA model's ability to recognize patterns and relationships beyond numerical proximity enhances clustering results. This contextual awareness can lead to more meaningful and clinically relevant clusters. 

The LLAMA 3.1 8B model identified five distinct patient clusters with statistically significant differences in demographic, clinical, and socioeconomic variables. \textbf{Cluster 1} consisted of the youngest patients with the highest in-hospital mortality rate. They exhibited poorer nutritional status and more severe clinical symptoms, such as severe respiratory distress and coma. These findings align with previous studies indicating that younger children with malnutrition are at higher risk of adverse outcomes\cite{black2013undernutrition, pelletier1995effects}. \textbf{Cluster 2} included older children with better nutritional indicators but a high incidence of malaria-positive cases. This suggests that while nutritional status is better, exposure to malaria remains a significant health concern, consistent with epidemiological data from malaria-endemic regions \cite{who2020malaria}. \textbf{Cluster 3} featured young children with the poorest nutritional indicators and higher prevalence of health issues, such as prolonged diarrhea and rash. Maternal factors, including lower education levels and higher rates of premature births, may contribute to the health challenges in this cluster, echoing findings that maternal education significantly impacts child health outcomes \cite{caldwell1979education, cleland1988maternal, victora2008adult}. \textbf{Cluster 0} represented children with stable clinical presentations. The high usage of standard antibiotics and better access to municipal water indicate favorable socioeconomic conditions, which are known to correlate with improved health outcomes \cite{victora2008adult}. \textbf{Cluster 4} was similar to Cluster 0 but exhibited specific clinical concerns, such as higher incidences of changes in urine color and potential HIV exposure. This underscores the importance of monitoring for renal conditions and addressing HIV-related healthcare needs \cite{gibb2003decline}.

These results suggest that LLM-based clustering not only simplifies the analysis of complex healthcare data but also provides actionable insights for targeted interventions and resource allocation \cite{haines2004bridging}. Observed associations—such as those between maternal education and child health—underscore the importance of educational initiatives \cite{frost2005maternal}. However, high computational requirements and the “black-box” nature of transformer models may limit clinical transparency and acceptance \cite{doshi-velez2017towards}.

\subsection{Conclusion}
LLM-driven representation learning and clustering offer distinct advantages by effectively handling high-dimensional, complex data and capturing non-linear relationships. The resulting patient clusters provide valuable insights for personalized healthcare and resource distribution, although further work is needed to optimize model efficiency and interpretability.

% References as numbers
\makeatletter
\renewcommand{\@biblabel}[1]{\hfill #1.}
\makeatother

% unstr is used to keep citation order
\bibliographystyle{vancouver}
\bibliography{cor_ref}

\end{document}